\documentclass[pre,preprint]{revtex4-2}
\usepackage{amsmath}
\usepackage{graphicx}

\newcommand{\deltat}{\tilde{\delta}}
\newcommand{\koff}{k_{\textrm{off}}}
\newcommand{\kon}{k_{\textrm{on}}}

\begin{document}

\title{Effective Model of Loop Extrusion Predicts Chromosomal Domains}
\author{Martina Crippa}
\affiliation{Department of Physics, Universit\`a degli Studi di Milano, via Celoria 16, 20133 Milano, Italy}
\affiliation{Department of Applied Science and Technology, Politecnico di Torino, Corso Duca degli Abruzzi 24, 10129 Torino, Italy}
\author{Yinxiu Zhan}
\affiliation{Friedrich Miescher Institute for Biomedical Research, Maulbeerstrasse 66, 4058 
Basel, Switzerland}
\author{Guido Tiana}
\email{guido.tiana@unimi.it}
\affiliation{Department of Physics and Center for Complexity and Biosystems, Universit\`a degli Studi di Milano and INFN, via Celoria 16, 20133 Milano, Italy}
\date{\today}
\begin{abstract}
An active loop-extrusion mechanism is regarded as the main out--of--equilibrium mechanism responsible for the structuring of megabase-sized domains in chromosomes. We developed a model to study the dynamics of the chromosome fibre by solving the kinetic equations associated with the motion of the extruder. By averaging out the position of the extruder along the chain, we build an effective equilibrium model capable of reproducing experimental contact maps based solely on the positions of extrusion--blocking proteins. We assessed the quality of the effective model using numerical simulations of chromosomal segments and comparing the results with explicit-extruder models and experimental data.
\end{abstract}

\maketitle

\section{Introduction}
Chromosomes display a hierarchical structure of domains during cellular interphase \cite{Dixon2012,Zhan2017}. In mammals, the level of topological associating domains (TADs), at the mega--base scale, constitutes the most important level in the hierarchy for their role in controlling gene expression. The folding of TADs has been described at a molecular level by an active loop--extrusion mechanism \cite{Fudenberg2016a}, where a protein complex extrudes chromatin loops and it can be stopped by proteins bound to chromosome (for a review, see refs. \cite{Fudenberg2017c,Ghosh2020x}). 

\begin{figure}
    \centering
    \includegraphics[width=8cm]{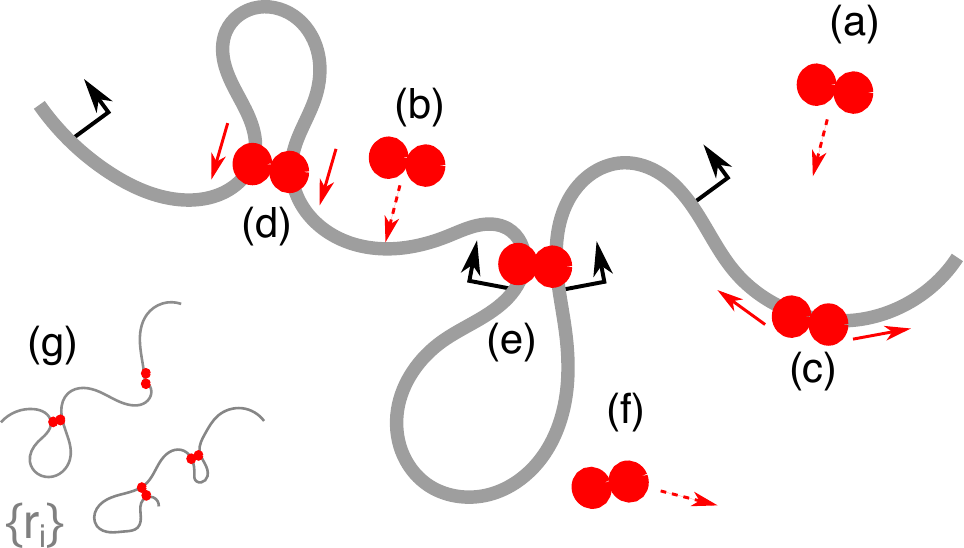}
    \caption{A sketch of the loop--extrusion mechanism. Cohesin diffuses in the nucleus (a) and can be loaded onto the chromosome at a random position (b). From here, it starts to run on the chromosome (c), extruding its two strings and thus forming a loop (d). CTCF proteins bound to chromosomes towards the running cohesin (black arrows) can stop tits motion (e). At any point, cohesin has a probability to detach from the chromosome (f). (g) The position of cohesin along the chain define the interactions that contribute to determining the conformations $\{r_i\}$ of the chromosome.}
    \label{fig:sketch}
\end{figure}

The cohesin protein complex has been suggested to extrude the chromatin fiber,  keeping close in space the two chromosomal segments at which it is bound at a given time (see Fig. \ref{fig:sketch}). The extrusion activity can be stopped by CTCF proteins bound to chromatin, thus stabilizing the contact between the CTCF-bound chromosomal regions. In fact, enrichment in CTCF has been observed in loci pivoting strong contacts \cite{Spencer2011}. Cells lacking either CTCF \cite{Wutz2017} or cohesin \cite{Rao2017a} display a reduced structuring of TADs. Recently, microscopy experiments using biochemically reconstructed systems showed that cohesin can extrude chromatin in an ATP--dependent way \cite{Davidson2019,Kim2019}.

An interesting feature of CTCF is that it is directional, in the sense that it can bind asymmetrically to chromatin in both directions and can stop efficiently cohesin only if it oriented towards it, but not those oriented opposite to it \cite{Sanborn2015}. This directionality arises because CTCF is not simply a barrier to the motion of cohesin, but interacts with it in a specific way. CTCF binds to DNA in a non-palindromic way \cite{Yin2017a} and a segment in its terminal segment interacts strongly with a specific domain of cohesin \cite{Li2020a}. The directionality of CTCF seems to be at the basis of "corner peaks" observed in contact maps of mammalian cells at the scale of 100 kbp \cite{Rao2014,Fudenberg2016a}.

Several polymeric models have been employed to describe the conformation and the dynamics of chromatin at the length scale of TADs \cite{Tiana2018}. They usually describe the chromosomal segments as a chain of beads interacting with some contact potential. Simulations of polymer chains including an additional degree of freedom that specify the position of extruders along the chain were shown to produce contact maps which are qualitatively similar to the experimental ones \cite{Fudenberg2016a,Sanborn2015,Nuebler2018}. Polymeric simulations of a diffusing extruder produced realistic contact maps also without energy consumption \cite{Brackley2017,Brackley2018}, even if experimental data suggest that ATP hydrolysis is a key ingredient for extrusion  \cite{Davidson2019,Kim2019}. A mechanistic model for ATP--dependent translocation of cohesin is described in ref. \cite{Marko2019}.

In the present work we studied the active, out--of--equilibrium dynamics of cohesin along the chromatin fiber and we built an effective model in which the position of cohesin along the chain is averaged out. In this way, we obtained a polymer model controlled by an effective potential whose equilibrium state reflect the distribution of conformations in cellular nuclei.

The reason why to build such an effective model is two--fold. On one side, it can be helpful to better understand the physics that controls the conformational properties of chromosomes. In fact, the reduction of the degrees of freedom maps chromosomes into systems that can be studied by standard polymer theory. On the other side, it can be a predictive tool to generate contact maps of chromosomal regions based on the position of CTCF, in a computationally more efficient way than explicit-extruder models.

\section{Motion of the extruder}
\label{sect:extruder}

Some of the numerical parameters that are necessary to build the model are known, see Appendix \ref{app:num}. In particular, the diffusion coefficient of cohesin in the nucleoplasm is much larger than that of chromatin loci on the TAD length scale, suggesting that one can assume cohesin to be well-mixed in cellular nucleus. Moreover, the time scale associated with extrusion is slightly smaller than that associated with the motion of the polymer chain on the TAD length scale. Even if this difference is marginal, we tested the assumption that the distribution of cohesin along the chain can be regarded as stationary. We compared the results of the effective model with both the experimental data and polymer model in which cohesin in simulated explicitly, thus without making in this case any assumption on its probability distribution.

Let's assume that the extruder can only walk towards the ends of the chain, that it walks with constant rate in a fixed direction and that it cannot overcome a CTCF molecule. Let's define the binary quantities $\sigma^+_i$ and $\sigma^-_i$ that assume the values $1$ if site $i$ contains a CTCF molecule oriented forward and backward, respectively. We also define 
\begin{equation}
\deltat^\pm_i\equiv  1-\delta_{\sigma^{\pm}_i, 1}
\end{equation}
that assumes the value 0 in the sites with a CTCF molecule oriented in the specified direction and thus it is able to stop the motion of the extruder in that direction; it takes the value 1 otherwise.

The rate equation that describes the amount $p_{i,j}(t)$ of extruder linking sites $i$ and $j$ of the chromosomes is
\begin{align}
&\frac{dp_{i,j}}{dt}=k_{\textrm{on}}\delta_{|i-j|,1}-\koff p_{i,j}+k\deltat^-_{i+1} p_{i+1,j}- \label{eq:first}\\
&-k\deltat^-_i p_{i,j}+k\deltat^+_{j-1} p_{i,j-1}-k\deltat^-_j p_{i,j}, \nonumber
\end{align}
where $k_{\textrm{on}}$ is the loading rate of the extruder on the chromosome, $\koff$ the detachment rate and $k$ the advancement rate.  The stationary distribution can be obtained setting to zero the time derivative for every $i$ and $j$, that is
\begin{equation}
p_{ij}=\frac{ k_{\textrm{on}}\delta_{|i-j|,1} + k\deltat^-_{i+1} p_{i+1,j} + k \deltat^+_{j-1} p_{i,j-1} }{ \koff + k( \deltat^-_i+ \deltat^+_j) }.
\label{eq:stat}
\end{equation}
This equation can be solved recursively, exploiting the fact that $p_{ij}$ depends only on the probabilities $p_{kl}$ such that $i<k<l<j$.

An important approximation that we implicitly did in Eq. (\ref{eq:first}) is that multiple extruders do not interact with each other by excluded volume when they walk on the chromosome.

\begin{figure}
    \centering
    \includegraphics[width=8cm]{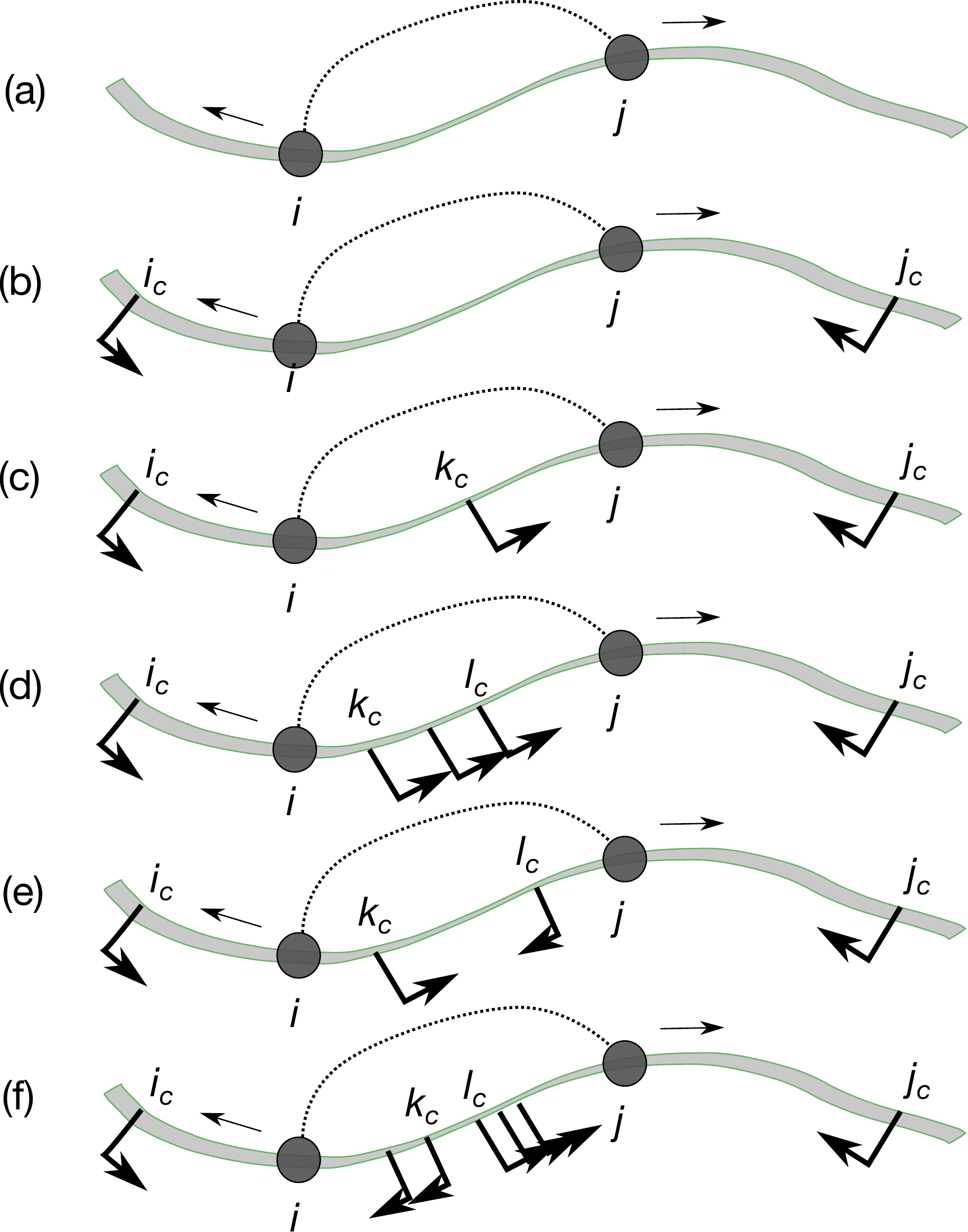}
    \caption{A sketch of the different ways in which cohesin can run, according to the position of CTCF. The pairs of bead indicate the position of cohesin at the two chromosomal sites it encloses.  (a) A chromosomal segment where the extruder can move freely. (b) The case in which the extruder is constrained by two convergent CTCF molecules (i.e., $\sigma^+_{i_c}=1$ and $\sigma^-_{j_c}=1$). (c) The case in which a further CTCF molecule prevents the motion of the extruder in one direction (i.e., $\sigma^+_{i_c}=1$, $\sigma^+_{k_c}=1$  and $\sigma^-_{j_c}=1$). (d) Is the case similar to the previous one, with multiple aligned CTCF. (e) The case with convergent CTCF in between. (f) The case of several divergent CTCF in between, the inner being at positions $k_c$ and $l_c$, respectively.}
    \label{fig:extruder}
\end{figure}

\subsection{Chromosome without CTCF}

The simplest case is that in which the extruder can walk freely on the chromosome in absence of CTCF, as described in Fig. \ref{fig:extruder}(a).

In this case, Eq. (\ref{eq:stat}) becomes
\begin{equation}
p_{i,j}=\frac{ k_{\textrm{on}}\delta_{|i-j|,1} + kp_{i+1,j} + kp_{i,j-1} }{ \koff + 2k }.
\label{eq:stat_noctcf}
\end{equation}
Starting from the case $i,i+1$ at which the extruder can bind, one can write iteratively
\begin{align}
p_{i,i+1}&=\frac{ k_{\textrm{on}} }{ \koff + 2k }\equiv p_{0}, \nonumber\\
p_{i,i+2}&=\frac{ kp_{i+1,i+2} + kp_{i,i+1} }{ \koff + 2k }=\frac{ 2k }{ \koff + 2k }p_0,\nonumber\\
p_{i,i+3}&=\frac{ kp_{i+1,i+3} + kp_{i,j+2} }{ \koff + 2k }=\left(\frac{ 2k }{ \koff + 2k }\right)^2 p_0\nonumber\\
&...\nonumber \\
p_{i,j}&=\left(\frac{ 2k }{ \koff + 2k }\right)^{j-i-1}p_0,
\label{eq:noCTCF}
\end{align}
where use is made of the translational invariance $p_{i+n,j+n}=p_{ij}$ and the boundary condition $p_{i,i}=0$.

\subsection{Contacts between sites within convergent CTCF}

Consider a chromosome segment bordered by convergent CTCF at sites $i_c$ and $j_c$, as in Fig. \ref{fig:extruder}(b). The value of $p_{ij}$ with $i_c<i<j<j_c$ depends only on the amount of extruder in the interval from $i$ to $j$, so for $i_c<i<j<j_c$ Eq. (\ref{eq:noCTCF}) still holds.

Equation (\ref{eq:stat}) can now be written as
\begin{equation}
p_{ij}=\frac{ k_{\textrm{on}}\delta_{|i-j|,1} + k(1-\delta_{i+1,i_c})p_{i+1,j} + k(1-\delta_{j-1,j_c})p_{i,j-1} }{ \koff + k(2-\delta_{i,i_c}- \delta_{j,j_c}) }.
\label{eq:stat1}
\end{equation}

The amount of extruder in sites containing a CTCF molecule can be found from Eqs. (\ref{eq:stat1}) and (\ref{eq:noCTCF}). For example, the term 
\begin{equation}
p_{i,j_c}=\frac{ kp_{i+1,j_c} + kp_{i,j_c-1} }{ \koff + k },
\label{eq:cctcf}
\end{equation}
where $p_{i,j_c-1}$ is that of Eq. (\ref{eq:noCTCF}) and we iterate on $p_{i+1,j_c}$. We get from Eqs. (\ref{eq:cctcf}) and (\ref{eq:stat1})
\begin{align}
p_{j_c-1,j_c}&=\frac{k_{on}}{\koff+k} \nonumber\\
p_{j_c-2,j_c}&=\frac{k}{\koff+k}\left[\frac{k_{on}}{\koff+k}+p_0\right]\nonumber\\
p_{j_c-3,j_c}&=\frac{k_{on}}{\koff+k}\left(\frac{k}{\koff+k}\right)^2 + \left(\frac{k}{\koff+k}\right)^2 p_0+\nonumber\\
+&\frac{k_{on}}{\koff+k}\left(\frac{k}{\koff+2k}\right)^2 p_0 \nonumber\\
p_{j_c-n,j_c}&=\frac{k_{on}}{\koff+k}\left(\frac{k}{\koff+k}\right)^{n-1}+\nonumber\\
&+\sum_{l=1}^{n-1} \left(\frac{k}{\koff+k}\right)^{n-l}\left(\frac{k}{\koff+2k}\right)^{l-1}p_0.
\end{align}
The general form of $p_{j_c-n,j_c}$ contains a geometric sum that gives
\begin{align}
&p_{j_c-n,j_c}=\frac{k_{on}}{\koff+k}\left(\frac{k}{\koff+k}\right)^{n-1}+\nonumber\\
&+\frac{\koff+2k}{k}\left(\frac{k}{\koff+k}\right)^{n-1}p_0-\left(\frac{k}{\koff+2k}\right)^{n-2}p_0
\label{eq:conv-1}
\end{align}
By symmetry, the same expression is valid for $p_{i_c,i_c+n}$. The probability associated with both CTCF sites obeys by Eq. (\ref{eq:stat1}) the relation
\begin{equation}
p_{i_c,j_c}=\frac{k p_{i_c+1,j_c} + k p_{i_c,j_c-1}}{\koff},
\label{eq:conv1}
\end{equation}
that can be evaluated substituting Eq. (\ref{eq:conv-1}) in it.

\subsection{Contacts across a CTCF site}
\label{sect:across}

Consider a segment from $i_c$ to $j_c$ closed by convergent CTCF molecules, with a further CTCF molecule at position $k_c$ with $i_c<k_c<j_c$ and, for instance, directed upward (i.e., $\sigma^+_{k_c}=1$), as in Fig. \ref{fig:extruder}(c).

Pairs of sites on the same side with respect to $k_c$ displays the same probabilities as described above, that is Eqs. (\ref{eq:stat_noctcf}), (\ref{eq:conv-1}) and (\ref{eq:conv1}). Pairs interspersed by CTCF molecules, i.e. $i<k_c<j$, are affected by the fact that the two sites cannot be reached evenly from extruders coming from all parts of the segment $(i,j)$.

Let's use again an iterative approach, starting from
\begin{equation}
p_{k_c-1,k_c}=p_0. 
\label{eq:p0across}
\end{equation}
The probabilities involoving site $k_c+1$ obey
\begin{align}
&p_{k_c-n,k_c+1}=\frac{k}{\koff+2k}p_{k_c-n+1,k_c+1}= \nonumber\\
&=\left(2^n-1\right)\left(\frac{k}{\koff+2k}\right)^np_0.
\label{eq:pi0}
\end{align}
Similarly, those involving site $k_c-1$ are given by Eq. (\ref{eq:noCTCF}),
\begin{equation}
p_{k_c-1,k_c+m}=\left(\frac{k}{\koff+2k}\right)^m p_0.
\label{eq:pi1}
\end{equation}
For any pair of sites across $k_c$, the probability obeys the iterative relation
\begin{equation}
p_{k_c-n,k_c+m}=\frac{k}{\koff+2k}\left(p_{k_c-n+1,k_c+m} + p_{k_c-n,k_c+m-1} \right).
\label{eq:iter1}
\end{equation}
One can look for solutions in the form
\begin{equation}
p_{k_c-n,k_c+m}=a_{n,m}\left(\frac{k}{\koff+2k}\right)^{n+m-1} p_0,
\end{equation}
that, substituted in Eq. (\ref{eq:iter1}), gives the iterative relation
\begin{equation}
a_{n,m}=\frac{k}{\koff+2k} (a_{n-1,m}+a_{n,m-1}),
\label{eq:iter}
\end{equation}
starting from $a_{n,1}=2^n-1$ (Eq. \ref{eq:pi0}) and $a_{1,m}=1$ (Eq.  \ref{eq:pi1}).

Solving the iterative problem making use of a bivariate generating function (see Appendix \ref{app:a}), one obtains
\begin{align}
p_{k_c-n,k_c+m}&= {{n+m-1}\choose{m}}\,_2F_1(1,1-n,1+m,-1)
\cdot \nonumber\\
\cdot&\left(\frac{k}{\koff+2k}\right)^{m+n-1}p_0,
\label{eq:pnm}
\end{align}
where $_2F_1$ is the Gaussian hypergeometric function.

\subsection{Contacts across several CTCF sites}

Consider now the case of a pair of sites $i$ and $j$ separated by more than a CTCF molecule, with various orientations, like in Figs. \ref{fig:extruder}(d)-(e). If both orientations are present, like in Fig. \ref{fig:extruder}(e), then $p_{i,j}=0$ because no extruder can bind to any pair of sites $q,q+1$ with $i<q<j$ and reach sites $i$ and $j$.

For sites  $i$ and $j$ separated by two CTCF sites (at positions $k_c$ and $l_c$) with the same alignment, as in Fig. \ref{fig:extruder}(d) one can follow the same strategy as that of Sect. \ref{sect:across}.

Analogously to Eq. (\ref{eq:p0across}), the starting point is the probability $p_{k_c-1,l_c}$ that in the present case is given by Eq. (\ref{eq:pi1}) because sites $k_c-1$ and $l_c$ fall in the case of Fig. \ref{fig:extruder}(c), that is
\begin{equation}
p_{k_c-1,l_c}= \left(\frac{k}{\koff+2k}\right)^{l_c-k_c}p_0.
\label{eq:it0many}
\end{equation}
From here, an iterative relation analogous to Eq. (\ref{eq:iter1}) holds, that is
\begin{equation}
p_{k_c-n,l_c+m}=\frac{k}{\koff+2k}\left(p_{k_c-n+1,l_c+m} + p_{k_c-n,l_c+m-1} \right),
\label{eq:iter2}
\end{equation}
whose solution is the same as that of Eq. (\ref{eq:pnm}), 
\begin{align}
p_{k_c-n,l_c+m}&={{n+m-1}\choose{m}}\,_2F_1(1,1-n.1+m,-1)\cdot\nonumber\\ 
\cdot&\left(\frac{k}{\koff+2k}\right)^{m+n+l_c-k_c-1}p_0,
\label{eq:pnm2}
\end{align}
with the difference that the iterative propagation is applied to Eq. (\ref{eq:it0many}) instead that to $p_0$ only. 

This solution can be easily extended to the case in which between the two sites of interest there is an arbitrary sequence $\{k_{c1},k_{c2},...,k_{cN}\}$ of CTCF sites aligned in the same direction. In this case, one can apply the propagator of Eq. (\ref{eq:pnm2}) to $p_{k_{c1,c(N-1)}}$ obtaining
\begin{align}
p_{k_{c1}-n,k_{cN}+m}&= {{n+m-1} \choose{m}}  \,_2F_1(1,1-n.1+m,-1)\cdot\nonumber\\
\cdot&\left(\frac{k}{\koff+2k}\right)^{m+n+cN-c1-1}p_0,
\label{eq:pnmN}
\end{align}
thanks to the fact that $_2F_1(1,0,1+m,-1)=1$.

The most problematic case is that of two sites $i$ and $j$ separated by diverging CTCF molecules, like in Fig. \ref{fig:extruder}(f). Calling $k_c$ and $l_c$, respectively, the inner sites, we know that
\begin{equation}
p_{k_c-1,l_c+1}=\frac{k}{\koff+2k}[p_{k_c-1,l_c}+p_{k_c,l_c+1}]    
\end{equation}
that can be easilly evaluated using Eq. (\ref{eq:pnm2}).
However, the exact solution of this case for generic values of $n$ and $m$ would require the summation of terms in the form of Eq. (\ref{eq:pnm2}), that we are not able to do. For this reason we resort to an approximation, writing
\begin{equation}
p_{k_c-n,l_c+m}=p_{k_c,l_c+m}\left(\frac{k}{\koff+k}\right)^n+  p_{k_c-n,l_c}\left(\frac{k}{\koff+k}\right)^m,
\label{eq:div}
\end{equation}
in which the probabilities at the right--hand side are given by Eq. (\ref{eq:pnm2}).
This corresponds to the assumption that the extruders that can reach sites $i$ and $j$ are only those that after reaching sites $k_c$ and $l_c+m$  walks $n$ steps on to reach $k_c+n$, and those that do the same thing from sites $k_c+n$ and $l_c$, making $m$ steps from the former. Equation (\ref{eq:div}) is exact for $n=m=1$ and is expected to underestimate the true probability for large $n$ and $m$, probability that is anyway low in this limit.
Moreover, under the same assumptions, the resulting probability does not change if multiple CTCF sites are aligned in the two directions, as in Fig. \ref{fig:extruder}(f).

\section{Effective model}

From the knowledge of the stationary distribution of extruder, we built an effective polymeric model in which the degrees of freedom of the extruder are averaged out. In other words, we started from a model which is surely out of equilibrium because the motion of cohesin does not obey the condition of detailed balance, we showed that the distribution of cohesin along the chain has a stationary distribution and investigated if there is an effective potential that displays that distribution at equilibrium, through Boltzmann statistics. Then, we used the parameters of this potential in the conformational space of the polymer, in connection with a realistic (although arbitrary) contact function that defines the spatial dependence of the potential.

Let's assume that the  number $\mu_{ij}$ of extruder molecules binding sites $i$ and $j$ of the chromosome can be written as an equilibrium state of an effective potential
\begin{equation}
P(\mu)=\frac{1}{Z_\mu}e^{-\sum_{ij}\epsilon_{ij}\mu_{ij}},
\label{eq:pareff}
\end{equation}
where $\epsilon_{ij}$ is a site-dependent effective energy.

Due to the rigid nature of the extruder, the conditional probability associated with a conformation $\{r_i\}$ of the system for any given state $\{\mu_{ij}\}$ of the extruder along the chain is
\begin{equation}
P(r|\mu)=\frac{1}{Z}\prod_{i<j}\delta(\Delta_{ij}(r)-\mu_{ij}) e^{-\beta U_0(r)},
\end{equation}
where $U_0(r)$ is an underlying potential describing excluded volume and other general features of the polymer, $\beta\equiv 1/kT$ and  $\Delta(r_{ij})$ is some  function that defines the approaching in space of two monomers. The functional form of $\Delta(r_{ij})$ is arbitrary and cannot be determined within the present theory; due to the physical features of the system, one can envisage some kind of short-range potential.

The marginal probability of a conformation, averaged over the extruder conformations is then
\begin{equation}
P(r)=\int d^{N^2}\mu \; P(r|\mu) \cdot P(\mu)
\end{equation}

Writing the delta function as $\delta(x)=\lim_{\kappa\to 0}\exp[x^2/2\kappa^2]$, the conformational probability is proportional to
\begin{equation}
\prod_{i<j}\int d\mu_{ij}\,\exp\left[ -\epsilon_{ij}\mu_{ij} -\frac{(\Delta_{ij}(r)-\mu_{ij})^2}{2\kappa^2}-\beta U_0(r)\right] \nonumber
\end{equation}
that is a Gaussian integral, that gives
\begin{equation}
P(r_i)=\exp\left[-\beta U_0(r)-\sum_{i<j}\epsilon_{i,j}\Delta_{i,j}(r) \right],
\end{equation}
where the term proportional to $\kappa\to 0$ has been dropped.

The effective potential has thus the form
\begin{equation}
U(r)=U_0(r)+k_BT\sum_{i<j}\epsilon_{i,j}\Delta_{i,j}(r)
\label{eq:utot}
\end{equation}
whose parameters can be found from Eq. (\ref{eq:pareff}) as
\begin{equation}
\epsilon_{i,j}=-k_BT\log P(\mu_{ij})-E_0
\label{eq:effective}
\end{equation}
and if the number of extruders is small (and thus $\mu_{ij}$ is essentially binary), the probabilities $p(\mu_{ij})$ can be regarded as proportional to the results $p_{i,j}$ of the rate equations calculated in Eqs. (\ref{eq:noCTCF}), (\ref{eq:conv-1}), (\ref{eq:conv1}), (\ref{eq:pnm}), (\ref{eq:pnmN}) and (\ref{eq:div}), according to the position and the orientation of CTCF. The arbitrary additive constant $E_0$ ($=-k_BT\log Z_\mu$) in the energies $\epsilon_{ij}$  must be set independently of the theory developed above (see below). Positive values of $\epsilon_{ij}$ are filtered to zero, because it is not realistic that the extruder induces a repulsive interaction in the polymer.

\section{Simulations with the effective model}
\label{sect:effectives}

To test the performance of the effective model, we performed molecular--dynamics simulations of chromosomal segments described by a chain of beads connected by springs and interacting with a potential $U=U_0+U_{\textrm{eff}}$ (cf. Eq. \ref{eq:utot}) given by a polymeric term
\begin{equation}
U_0=\frac{k_s}{2}(\Delta r_{ij}-a)^2+\epsilon_0\sum_{i<j}\left(\frac{2a^{6}}{\Delta r_{ij}^{6}}-\frac{a^{12}}{\Delta r_{ij}^{12}}\right),
\end{equation}
where $\Delta r_{ij}\equiv|r_i-r_j|$, and the effective potential
\begin{equation}
U_{\textrm{eff}}=\sum_{i<j}\epsilon_{i,j}\left(\frac{2a^{6}}{\Delta r_{ij}^{6}}-\frac{a^{12}}{\Delta r_{ij}^{12}}\right)
\end{equation}
representing the effect of the extruder. The rest distance $a$ of the harmonic spring sets the elementary scale of the system, that is $a=67\;nm$ corresponding to a resolution of $5\cdot 10^3\;bp$  (cf. appendix \ref{app:num}). The interaction range of the Lennard--Jones potentials is set to $a$ as well. All simulations are performed at room temperature, $k_BT=2.5\;kJ/mol$. The harmonic constant is set to allow $10\%$ fluctuations of the spring length, that is $k_s=10^2k_BT/a^2=250\; kJ/mol/a^2$. The value of $\epsilon_0$ is set to $-1.5\;kJ/mol$ so that simulations in absence of extruders ($U_{\textrm{eff}}=0$) display the polymer fragment at the $\theta$--point. This is the simplest choice assuming that there are no other active mechanisms besides cohesin extrusion. In this case, the chromosome fragment we simulated mimics a segment of a much larger polymeric system and thus should obey ideal--chain statistics, in accordance with Flory theorem \cite{Grosberg1994}. To be noted that experiments depleting cohesin display a contact probability that scales with the linear distance as a power law with exponent 1.2 \cite{Schwarzer2017}, which is not that of an ideal chain (1.5), suggesting that some other out-of-equilibrium mechanism is at work. Choosing to simulate the polymer at the $\theta$--point we neglect these other mechanisms.

\begin{figure}
    \centering
    \includegraphics[width=8cm]{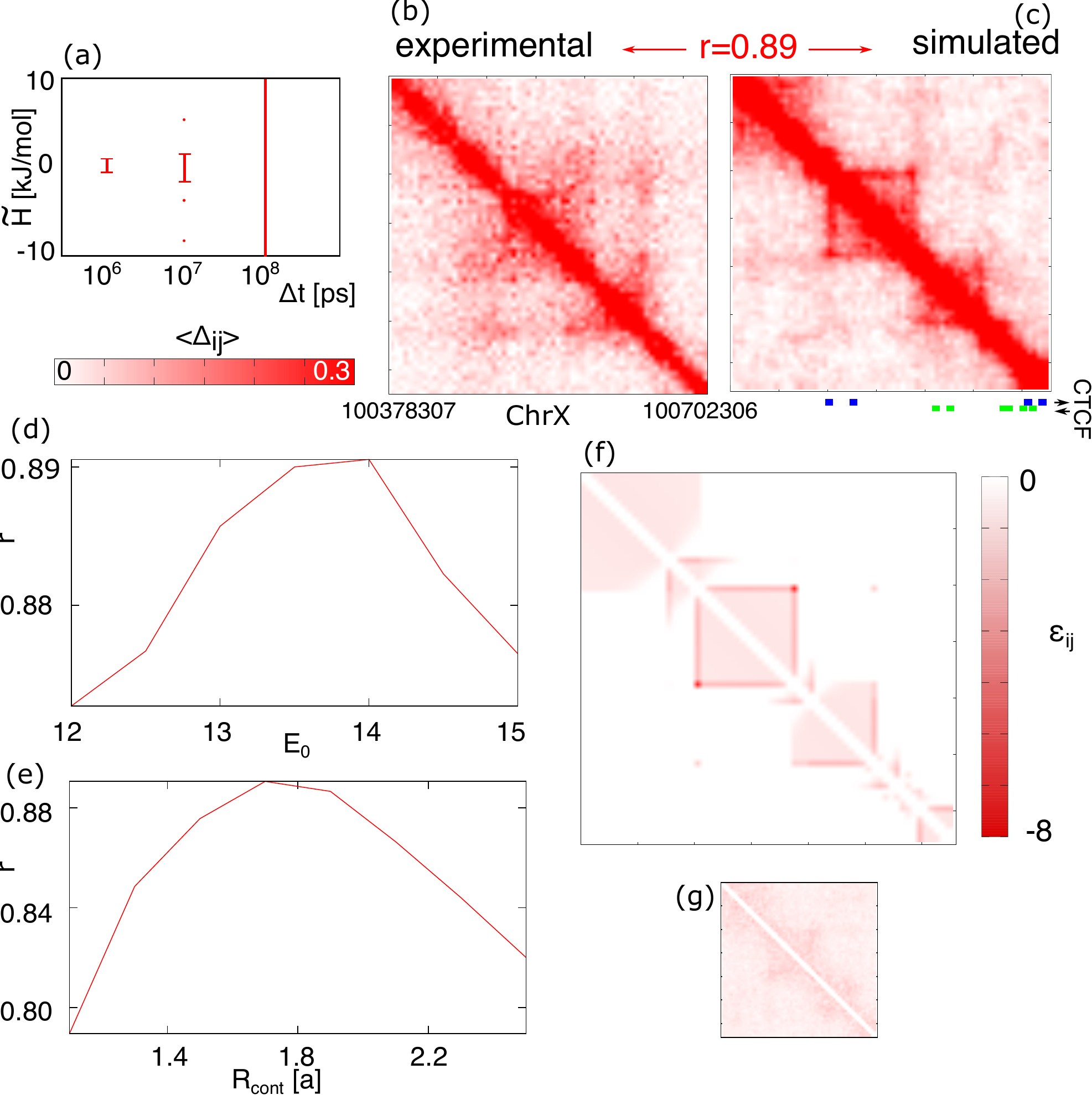}
    \caption{(a) The width of fluctuations of the effective energy $\tilde{H}$ as a function of the time step $\Delta t$ of the simulation, in ps. (b) The experimental Hi--C map of the Tsix region. (c) The contact map simulated with the effective model. The positions of CTCF in both orientations are indicated below the map. The correlation coefficient $r$ between simulated and experimental contact map as a function of the shift $E_0$ of the energy parameters (d) and of the distance $R_{cont}$ that defines contacts (e). (f) The interaction energy $\epsilon_{ij}$ between the beads. (g) The standard deviation between 10 simulations of 16 minutes each, plotted in the same color scale of the contact maps.}
    \label{fig:implicitTsix}
\end{figure}

Simulations are performed solving Langevin equations with Euler's integrator. For time steps $\Delta t$ much larger than $m/\gamma\approx10^{-4}ps$ (see Appendix \ref{app:num}) one can use the first--order overdamped version of Langevin equations. The fastest degree of freedom is that associated with the harmonic springs defining the chain, so we expect the time step to be smaller than the associated time scale, that is $\Delta t<\gamma/k_s\sim 10^{10}\;ps$. We evaluated the quality of the simulation quantifying to which extent it satisfies the principle of detailed balance. For this purpose, we calculated the quantity
\begin{equation}
{\tilde H} \equiv \frac{\Delta t}{4\gamma}\left( \frac{\partial U}{\partial r}\right)^2-\frac{r}{2}\frac{\partial U}{\partial r}  +U(r)
\end{equation}
that is a parameter (in energy units) which is strictly conserved if detailed balance is satisfied \cite{Bussi2007a}, in the present case by the Euler integrator. In simulations performed with $\Delta t\leq 10^7\;ps$ ${\tilde H}$ is conserved within an error of $k_BT$ (see Fig. \ref{fig:implicitTsix}(a) ). For larger values of $\Delta t$ this is no longer the case, and thus the simulations are no longer correct. We used $\Delta t= 10^7\;ps$ in the rest of the simulations. 

We first applied the effective model to a small region of mouse embryonic stem cells (of coordinates ChrX:100378307-100702306, the so--called Tsix TAD). The position and orientation of CTCF are taken from ref. \cite{Nora2017}. The map of the interaction energies $\epsilon_{ij}$ is disèlayed in Fig. \ref{fig:implicitTsix}(f); the main patterns displayed by the Hi--C map (cf. Fig.  \ref{fig:implicitTsix}(b)) are already apparent here. The average contact map $\overline{\Delta}_{ij}$ obtained from simulations of 16 minutes each is displayed in Fig. \ref{fig:implicitTsix}(c) and is compared with the experimental map obtained from Hi--C experiments (Fig. \ref{fig:implicitTsix}(b), \cite{Redolfi2019}). After this time span, contact maps are at convergence and one can expect that the system is equilibrated.  

In this procedure there are two free parameters that have to be set, namely the additive constant $E_0$ associated with the interaction energies (cf. Eq. (\ref{eq:effective})) and the cut--off distance $R_{cont}$ for two beads to be defined as in contact, distance that can hardly be obtained from a molecular insight of the crosslinking process at the basis of Hi--C maps,. These parameters are then obtained maximizing the Pearson's correlation function $r$ between all pairs of contacts in the simulated and in the experimental contact map. The correlation coefficient $r$ as a function of $E_0$ is displayed in Fig. \ref{fig:implicitTsix}(d) and display a maximum at $E_0=14$. The dependence on  $R_{cont}$ is displayed in Fig. \ref{fig:implicitTsix}(e) and is optimal at $R_{cont}=1.7a$. The optimal correlation obtained with these values is $r=0.89$. The simulated contact map displays the main features of the experimental Hi--C map, including two regions with high contact probability (cf panels b and c in Fig. \ref{fig:implicitTsix}). Moreover, the simulated map displays strong contacts in the initial part of the polymer which are not present in the experimental map; this is a region lacking of any CTCF molecule.

\begin{figure}
    \centering
    \includegraphics[width=6cm]{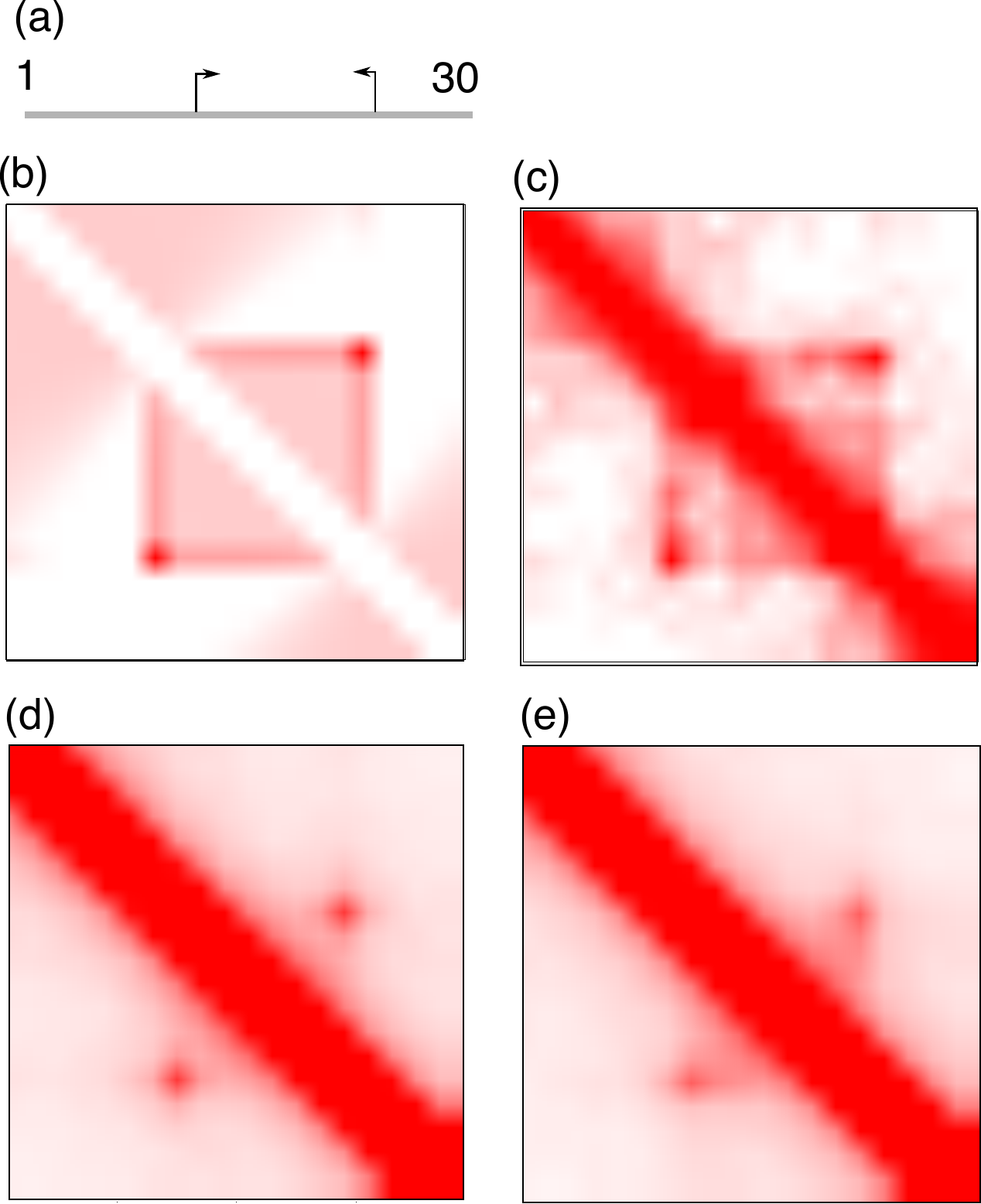}
    \caption{(a) A toy model with two convergent CTCF molecules. (b) The effective contact energy $\epsilon_{ij}$. (c) The result of a simulation with the effective model of 16 minutes with $E_0=13$. (d) The contact map resulting from the average of 30 explicit-extruder simulations of 16 minutes each in which extruders can overcome each other freely. (e) The result of simulations in which extruders cannot move into occupied sites. }
\label{fig:toymodel}
\end{figure}

To better understand the effective model, we simulated a toy model made of a 30-bead string with two convergent CTCF sites, as displayed in Fig. \ref{fig:toymodel}(a). The effective interactions $\epsilon_{ij}$ display a square of strongly interacting elements within the two CTCF sites, see Fig. \ref{fig:toymodel}(b); the borders of this square are even more interacting, as well as the corner where the extruder accumulates. In addition, there is a sort of 'border effect' due to extruders that bind to the CTCF--free ends of the chain. The simulation of the effective model produces a contact map which reflects essentially the interaction potential (cf. \ref{fig:toymodel}(c)). 

\section{Comparison with explicit extruder model}
\label{sect:explicit}

A relevant question we want to answer is how the the effective model performs with respect to a model in which the extruder is described explicitly \cite{Fudenberg2016a}. In fact, we do not expect that the effective model can reproduce all details of Hi--C maps, because loop extrusion is not the only mechanism at work. For example, it is known that the formation of compartments on the scale of the whole chromosome is not driven by loop extrusion but interacts with it at a smaller scale \cite{Schwarzer2017}. Our main goal is then to show that the effective model can reproduce Hi--C map with the same accuracy as the explicit--extruder model.

In the explicit model, we assumed that the  extruder is well--mixed around the polymer and it is always available for binding. It can bind to a pair of adjacent sites with rate $\kon$, each side  of the extruder can walk with rate $k$ and it can detach with rate $\koff$. The monomers linked by an extruder experience a harmonic force characterized by a harmonic constant $k_s$ and a rest distance $a$, that is the same force that guarantees the integrity of the polymer. We assumed that different bound extruders cannot overcome each other and they cannot overcome CTCF sites. The numerical parameters are given in Appendix \ref{app:num}.

\begin{figure}
    \centering
    \includegraphics[width=8cm]{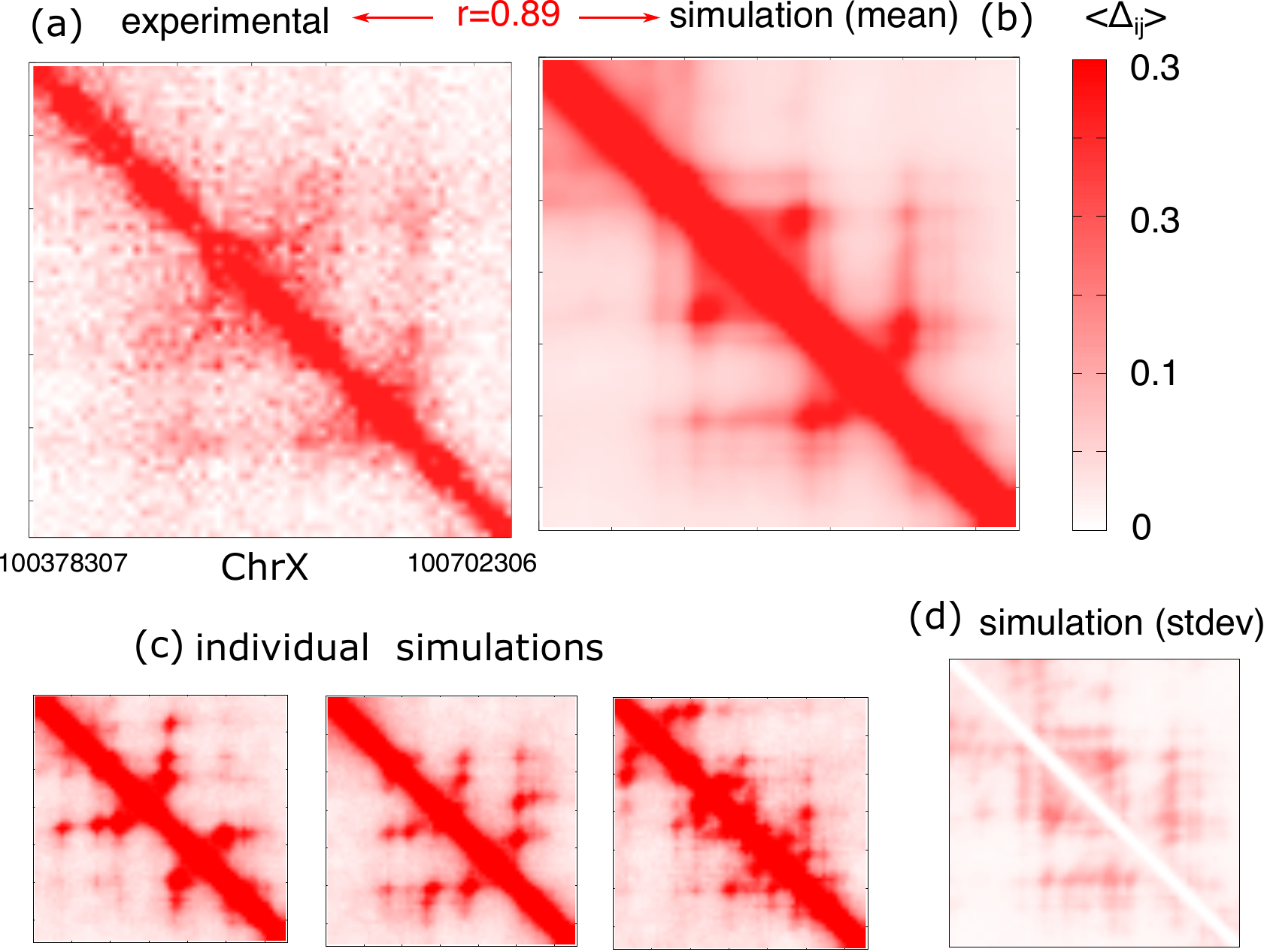}
    \caption{Results of the simulation of the Tsix TAD with explicit extruders. (a) The experimental map. (b) The mean contact function $\langle\Delta_{ij}\rangle$, averaged over 30 simulations (c) Examples of individual simulations that contribute to the average. (d) Their standard deviation.}
    \label{fig:explicit}
\end{figure}

The average contact map obtained from 30 simulations, calculated in the same way as those obtained with the effective model (cf. Sect. \ref{sect:effectives}), is displayed in Figs. \ref{fig:explicit}(a) and (b) for the Tsix domain. Contact maps seem to be at convergence. The correlation coefficient with the experimental map is $r=0.89$, which is identical to that of the effective model. 

The main difference between the explicit and the effective model is in the fluctuations around the average. In Fig.  \ref{fig:explicit}(c) we showed the result of three individual simulations and in Fig. \ref{fig:explicit}(d) the standard deviation associated with the  simulations. It is apparent that in explicit--extruder simulations the average map is given by the contribution of maps which are quite different from each other. In fact, the standard deviation is comparable with the average. This result is different than that of the effective model, in which the different simulations generate maps which are much more homogeneous (cf. Fig. \ref{fig:implicitTsix}(g) ).

In the case of the toy model of Fig. \ref{fig:toymodel}, explicit-extruder simulations produce contact maps in which corner peaks are more evident, and the overall domain is less clear. We also compared the results of simulations in which extuders are freely allowed to overcome each other (Fig. \ref{fig:toymodel}(d)) with simulations in which an extruder cannot occupy a site which is already occupied (Fig. \ref{fig:toymodel}(e)). The two maps are essentially identical, suggesting that the hypothesis done in connection with Eq. (\ref{eq:stat_noctcf}) is not critical.

\begin{figure}
    \centering
    \includegraphics[width=8cm]{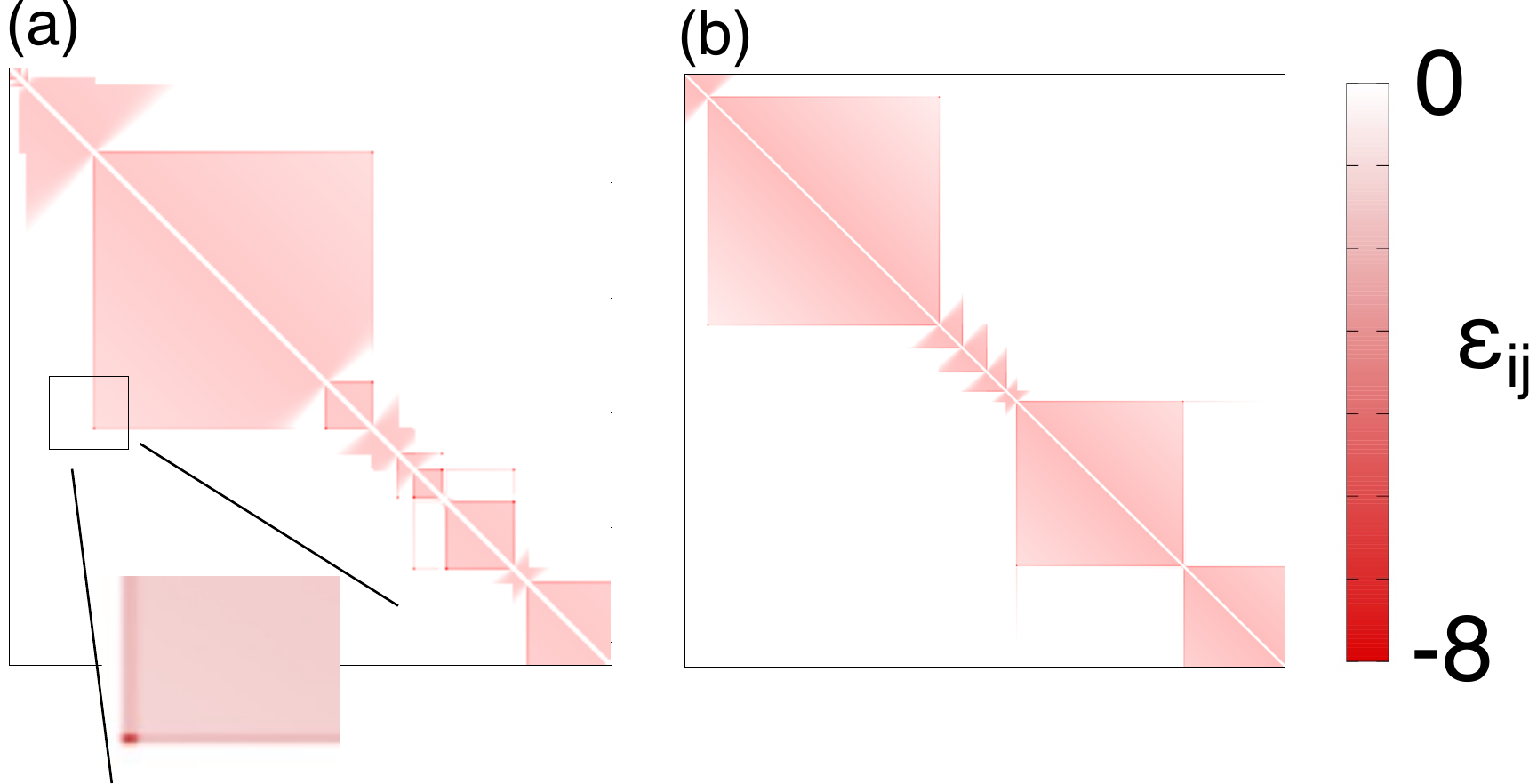}
    \caption{The interaction energies $\epsilon_{ij}$ between beads for (a) region chrX:102278307-103570000 and (b) region chrX:103578307-106170000. The inset is the zoom of the squared region.}
    \label{fig:potential}
\end{figure}

\begin{figure}
    \centering
    \includegraphics[width=8cm]{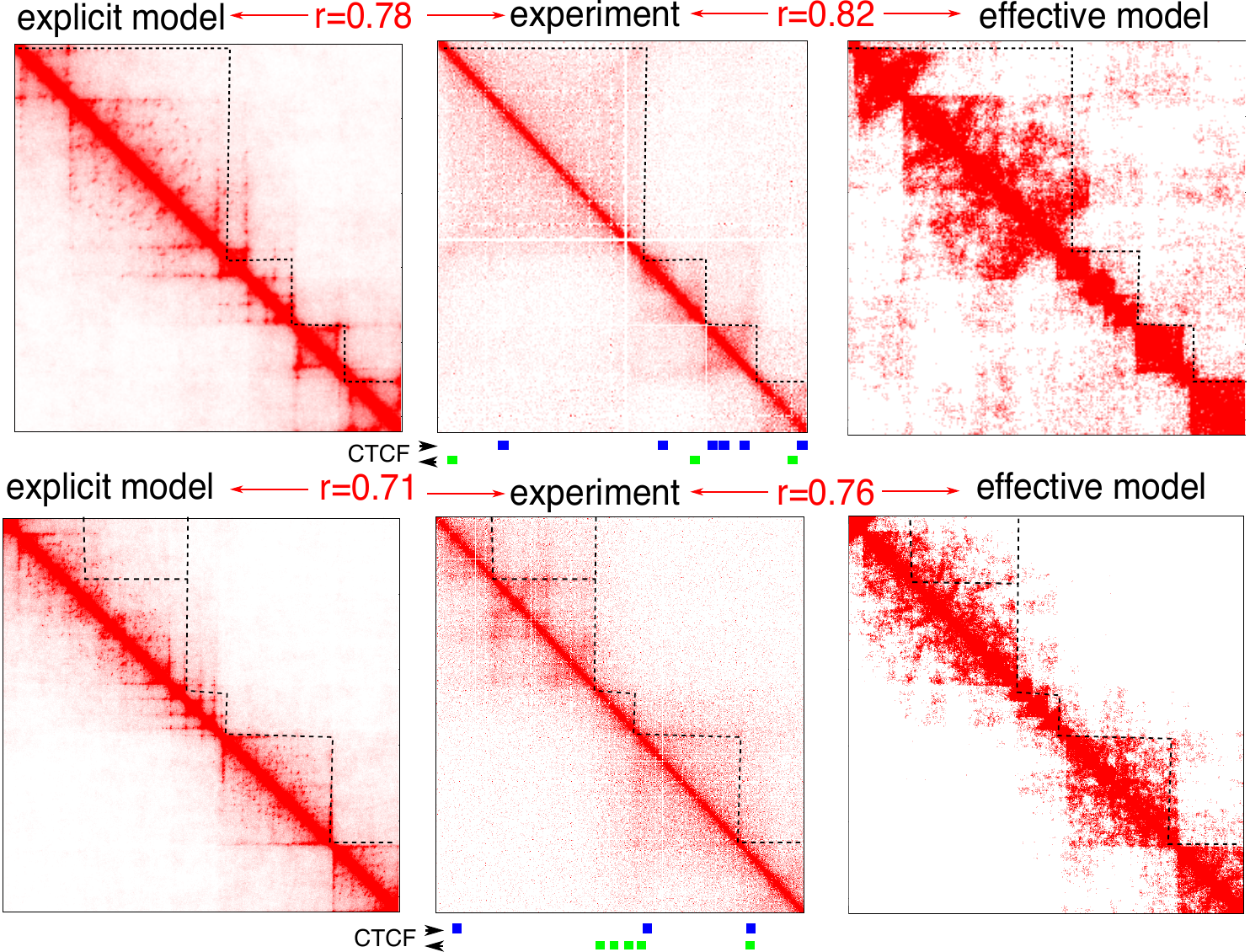}
    \caption{Comparison between the experimental data (central panels), the results of the explicit--extruder model (left panels) and those of the effective model (right panels). The upper panels are region chrX:102278307-103570000, the lower panel chrX:103578307-106170000. The dashed lines are a guide to the eye.}
    \label{fig:summa}
\end{figure}

We repeated similar calculations for other two regions of the chromosome X of mouse embryonic stem cells, of 1300 and 2600 kbp, respectively.  The energy maps $\epsilon_{ij}$ are displayed in Fig. \ref{fig:potential}. As in the case of Tsix, the energy maps contain most of the features displayed by the experimental maps. The comparison between the results of the explicit--extruder model, those of the effective model and the experimental map are displayed in Fig. \ref{fig:summa}. Also in these cases, the effective model ($r=0.82$ and $r=0.76$ for the two regions, respectively) performs similarly, if not better, than the explicit--extruder model ($r=0.78$ and $r=0.71$, respectively). 
The main features of the Hi--C maps are captured by both the explicit and the effective models. While small TADs are captured well, large TADs display in the models a finer structure which is not apparent in the experiment. Interestingly, the explicit and the effective models produce very similar maps (r=0.89 and 0.80, respectively), even in the patterns that are not in the experimental map.

The larger correlation of the effective model with the experimental data is presumably due to simulation statistics: while the explicit-extruder model requires to average the motion of the extruders over multiple simulations (30 in the case shown above), the effective model already contains implicitly this average.  The most apparent difference between the two models is that while the explict extruder generates maps whose elements are  spatially correlated with their neighbors, the maps obtained with the effective model display abrupt changes between neighboring elements. This is not unexpected, since the effective model, assuming a stationary distribution of the extruder along the chain, neglects correlations between consecutive sites associated with the motion of the extruder on short time scales.

A popular way of summarizing the information contained in contact maps of chromosomes is studying the average contact probability between sites as a function of their distance $|i-j|$ along the chain, that is usually a power law \cite{Lieberman-Aiden2009}.

The three sets of experimental data we studied display scaling coefficients $\beta=0.71$, $\beta=0.87$ and $0.77$, respectively. The simulations display a central region (around 10 beads, corresponding to 50 kbp) in which the contact probability is a power law with good exponents both in the case of the explicit--extruder (giving 0.74, 0.86 and  0.75, respectively) and the effective model (0.73, 0.86 and 0.78, respectively). In addition, all models display a bend at $|i-j|<5$, likely due to the coarse--graining of the model and an exponential cut-off due to finite--size effects.

This power--law dependence of the contact probability on the linear distance is not surprising in the light of the effective model, if this describes realistically the effective interactions between the beads of the polymer. In fact, both the interaction energy (cf. Eq. \ref{eq:effective}) and the polymer looping entropy display such a power--law dependence.

\begin{figure}
    \centering
    \includegraphics[width=7cm]{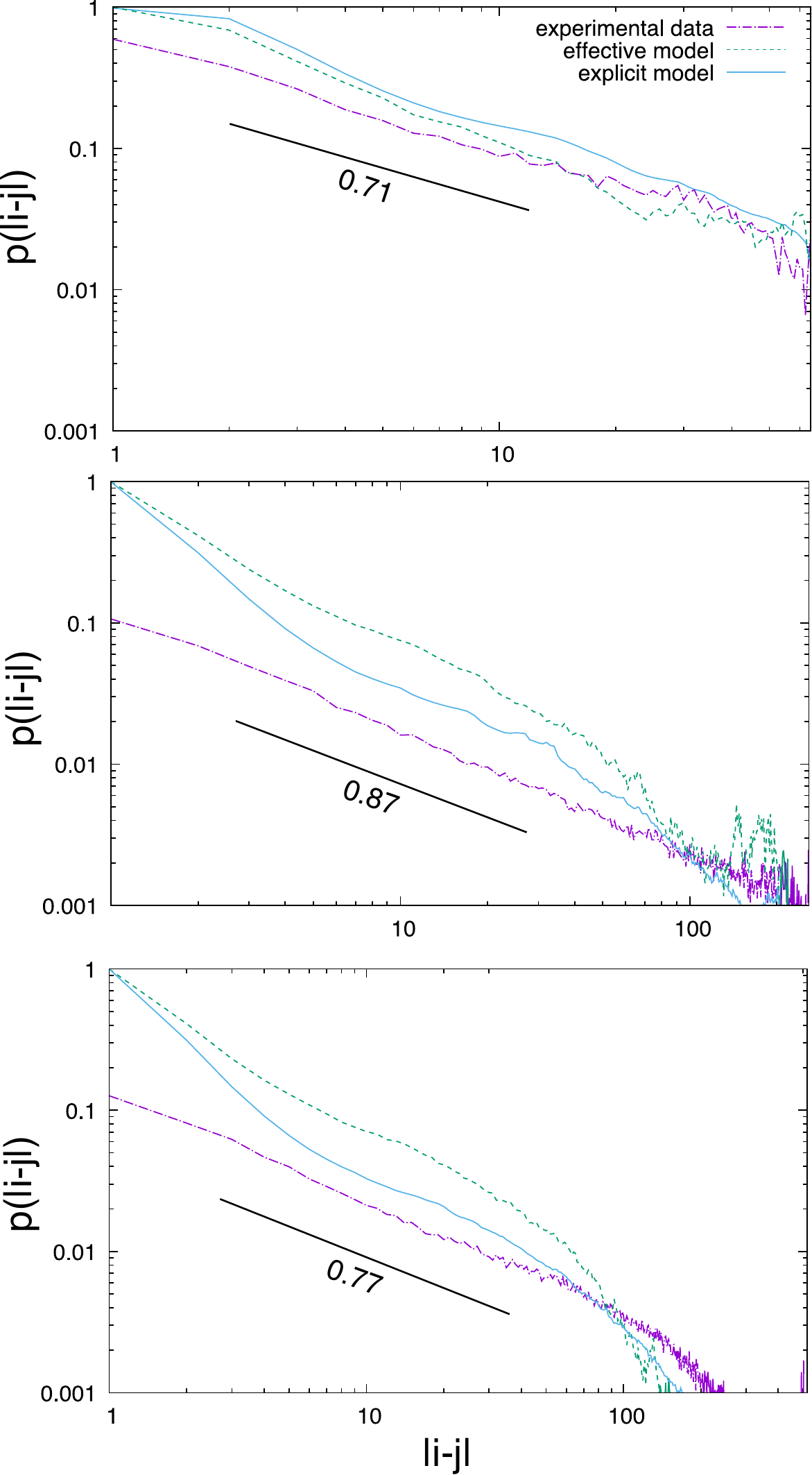}
    \caption{The contact probability between sites as a function of their distance along the chain, for the Tsix domain (upper panel), for region chrX:102278307-103570000 (middle panel) and region chrX:103578307-106170000 (lower panel).}
    \label{fig:scaling}
\end{figure}

\section{Conclusions}

We developed an effective model for the dynamics of chromosomes based on the assumption that the interactions that stabilize TADs are mediated by extruders running along the polymer consuming energy. The effective model is built in such a way that its equilibrium conformations approximate the conformations visited in the long run by the out--of--equilibrium extrusion mechanism.

We showed that simulations performed with the effective model produce average contact maps that are as similar to the experimental Hi--C map as those from an explicit extrusion mechanism. Even if they allow to detect TADs, the agreement with the experimental data for both kinds of model is still not perfect. This is not surprising because they are based on a minimal amount of information, that is the position of CTCF along the chain. There are indeed models \cite{Giorgetti2014,Marenduzzo2016,Jost2017} that produce contact maps closer to the experimental ones, but at the price of a larger amount of input information, being thus less predictive. Importantly, the present model can be improved by adding information from the experimental maps, such as the presence of compartments on a length scale larger than that of TADs \cite{Nuebler2018}.

The effective model in controlled by a free parameter ($E_0$) because interaction energies are defined but for an additive constant. This parameter cannot be determined by the position of CTCF but has to be tuned manually. Although the details of the simulated maps depend on this choice, the overall partitioning of the chromosome into domains seems quite robust with respect to it.

The effective model is based on averaging out the position of the extruder along the chain, so it is a mean--field approximation. Although this appears to be good enough to reproduce average maps, by definition it cannot account for fluctuations. Thus, the price to be payed to reduce the complexity in the description of the system is the loss of information about cell--to--cell variability.

Nonetheless, the maps that summarize the interaction energies $\epsilon_{ij}$ display the main patterns present in the experimental Hi--C maps, suggesting that polymeric entropy plays a limited role in shaping the architecture of chromosomes, at least at the scale of Mbp. 

Another strong approximation that we implemented is that the extruder cannot overcome CTCF sites. This approximation allowed us to obtain an analytical expression for the distribution of the extruder. However, it is known that CTCF is bound to its binding sites only for $\approx 50\%$ of time \cite{Hansen2017as}, resulting in an effective permeability of CTCF sites. Explicit simulations of the motion of phantom and of sterically-interacting extruders give similar results, at least in a simple toy system.

An approach analogous to ours was followed in ref. \cite{Brackley2017}, in which a Fokker--Planck equation for the binding probability of the extruder is solved in case of constant velocity, of pure diffusion and of diffusion in an effective potential that reflects the entropic cost of polymer looping. Only in the last case the binding probability displays a power-law scaling, reflecting the dependence of the entropy cost on the linear distance of the loop. However, recent experiments \cite{Davidson2019,Kim2019} indicate that cohesin uses ATP not only to bind/unbind but also to run on the chromosome. Since ATP hydrolysis rate in cohesin is approximately 2 s$^{-1}$ \cite{Davidson2019}, assuming that hydrolysis provides $\approx 30$ kJ/mol, the provided power is approximately 60 kJ/mol per second. On the other hand, cohesin runs at a rate of 2 kbp/s, making a loop of 4 kbp every second. Assuming a persistence length of the order of a kbp, the energy loss associated with the formation of the loop is at most $T\log 4\approx 3$ kJ/mol, which is one twentieth of the provided energy. Consequently, one does not expect entropy loss to be the main determinant of the distribution of cohesin on the fiber.

The computational gain offered by the effective model is quite consistent, not only because one has not to solve the equations for the extruder, but also because in a single simulation one describes in an effective way multiple trajectories. As an example, the simulation of the Tsix domain with the effective model takes of the order of 0.2 h per simulated minute per cpu core, to be compared with 6 h with the explicit model.

The effective model is useful not only to make simulations more efficient, but also to clarify the physics of chromatin. For example, in the light of the form of the potential developed for the effective model, it is not surprising that the contact probability scales with the linear distance as a power law. In fact, both the looping energy and the associated entropy loss scale as power laws. 

Te code to perform the simulations can be downloaded at https://github.com/martanit/LEAD

\appendix 
\section{Numerical parameters of the system}
\label{app:num}

 Experiments of fluorescence recovery after photobleaching indicate that the mean residence time of cohesin on the chromatin fiber is 13 minutes, corresponding to a detachment rate $\koff=1.3\cdot 10^{-3}$s$^{-1}$ \cite{Holzmann2019}. Total internal reflection microscopy of reconstructed cohesin--chromatin in a flow cell indicate that the stepping rate of cohesin is $k=10^3$ bp/s \cite{Davidson2019}. Fluorescence correlation spectroscopy experiments show that approximately $c=250,000$ copies of cohesin are present in human cells in G1 phase and that 64\% of them are bound to chromatin \cite{Holzmann2019}. The same order of magnitude but a smaller number ($c=109,000$) is obtained for mouse embryonic stem cells \cite{Cattoglio2019}. The binding rate $\kon'$ of cohesin on chromatin per base can be estimated from $\koff$ and from the fraction of bound molecules, that is
\begin{equation}
\kon'=\koff\frac{V_n}{N}\frac{c_b}{c-c_b},.
\end{equation}
where $V_n$ is the nuclear volume, $N$ is the total number of base pairs, $c_b$ is the number of bound cohesin molecules and $c$ is the total number of cohesin molecules. Using $V_n=500$ $\mu m^3$, $N=3\cdot 10^9$, $c=2.5\cdot 10^5$ and $c_b=1.6\cdot 10^5$ \cite{Holzmann2019} one obtains $\kon'=3.9\cdot 10^{-13}$ $nm^2/(ps\cdot bp)$. 

The (effective) diffusion coefficient of the chromatin fiber measured by live--cell imaging is $D_{ch}=3\cdot 10^{-3}$ $\mu m^2/s$ \cite{Tiana2016}. The diffusion coefficient of cohesin can be estimated by Stoke's law, using a hydrodynamic radius of $R=8.5$ nm \cite{Weitzer2003} and a viscosity for the nucleoplasm of $\eta=1.5$ cP \cite{Liang2009}. One obtains for cohesin $D_{co}=18$ $\mu m^2/s$, which is five orders of magnitude larger than that of chromatin, justifying the well--mixed hypothesis.

The time scale $\tau_{ext}$ associated with extrusion on the TAD length scale (i.e., $L\sim 10^6\; bp$ extruded by $n_{ext}\sim 30$ cohesin molecules) is $\tau_{ext}\sim L/(k\,n_{ext})\sim 30\;s$. The time scale associated with the motion of the chain is $\tau\sim L^2/D_{ch}\sim 300$s (using for typical TADs $L\sim 100\;nm$ \cite{Boettiger2016}).

The Hi--C maps we used as reference have a resolution of $5\cdot 10^3$ bp, an thus we used this as elementary unit of the model. From the density obtained from ref. \cite{Giorgetti2014}, this corresponds to $a=67\;nm$. We used this quantity as elementary length scale for the model. The friction constant of the polymer can be obtained from Einstein's equation $\gamma=D_{ch}/k_BT$ and, in terms of the length scale $a$, at room temperature is $\gamma=4\cdot 10^{12}\;kJ\, ps/mol\, a^2$. The stepping rate of cohesin is $k=2\cdot 10^{-13}\;a/ps$. The loading rate used for simulations of $N$ monomers is $\kon=N\kon'$.

Assuming the mass density typical of biomolecules, $1\;g/cm^3$, the mass of a monomer is of the order of $10^{-22}\;Kg$.

\section{Solution of the recursive equation}
\label{app:a}

Let's define $b_{n,m}=a_{n+1,m+1}$. Eq. (\ref{eq:itera}) can be written as
\begin{equation}
b_{n,m}=b_{n-1,m}+b_{n,m-1},
\label{eq:itera}
\end{equation}
that can be solved iteratively starting from $b_{n,0}=2^{n+1}-1$ and $b_{0,m}=1$.

Let's define the bivariate generating function
\begin{equation}
f(x,y)=\sum_{n,m=0}^\infty b_{n,m}x^n y^m
\end{equation}
Separating the terms $m,n=0$ one obtains
\begin{align}
f(x,y)&= \sum_{n=0} (2^{n+1}-1)x^n + \sum_{m>0} y^n + \sum_{n,m>0} b_{n,m}x^n y^m \nonumber\\
=&\frac{1}{1-x}+\frac{y}{1-y} + \sum_{n,m>0} b_{n,m}x^n y^m.
\end{align}
Substituting Eq. (\ref{eq:itera}) and renaming the indexes,
\begin{align}
f(x,y)&=\frac{1}{(1-x)(1-2x)}+\nonumber\\
+&\frac{y}{1-y} + \sum_{n,m>0} [b_{n-1,m}+b_{n,m-1}] x^n y^m =\nonumber\\
=& \frac{1}{(1-x)(1-2x)}+\frac{y}{1-y} + x\sum_{n=0,m>0}b_{n,m}x^ny^m-\nonumber\\
+& y\sum_{n>0,m=0}b_{n,m}x^ny^m =\nonumber\\
=& \frac{1}{(1-x)(1-2x)}+\frac{y}{1-y} + xf(x,y)-\nonumber\\
-&\frac{x}{(1-2x)(1-x)}+yf(x,y)-\frac{y}{1-y}=\nonumber\\
=& \frac{1}{(1-2x)}+
xf(x,y)+yf(x,y)
\end{align}
Thus,
\begin{equation}
f(x,y)=\frac{1}{(1-x-y)(1-2x)},    
\end{equation}
whose series expansion is
\begin{equation}
b_{n,m}=2^{\scriptscriptstyle n+m+1}-\frac{\Gamma(n+m+2)_2F_1(1,n+m+2,n+2,1/2)}{2\Gamma(n+2)\Gamma(m+1)}
\end{equation}
where $_2F_1$ is the Gaussian hypergeometric function.
This expression can be simplified to
\begin{equation}
b_{n,m}={{n+m+1}\choose{m+1}}\,_2F_1(1,-n.2+m,-1)
\end{equation}
and thus
\begin{equation}
a_{n,m}={{n+m-1}\choose{m}}\,_2F_1(1,1-n.1+m,-1)
\end{equation}

\end{document}